\documentclass[aps,reprint,superscriptaddress]{revtex4-1}
\usepackage{amsfonts}
\usepackage{graphicx}
\usepackage{dcolumn}
\usepackage{bm}
\usepackage{amsmath}
\usepackage{amssymb}
\usepackage{siunitx}

\begin{document}

\title{Optically and Electrically Tunable Dirac Points and Zitterbewegung in
Graphene-Based Photonic Superlattices}
\author{Hanying Deng}
\affiliation{Department of Physics and Astronomy, Shanghai Jiao Tong University, Shanghai 200240,
China}
\affiliation{Key Laboratory for Laser Plasma (Ministry of Education), IFSA Collaborative Innovation Center, Shanghai Jiao Tong University, Shanghai 200204, China}

\author{Fangwei Ye}\email[]{fangweiye@sjtu.edu.cn}
\affiliation{Department of Physics and Astronomy, Shanghai Jiao Tong University, Shanghai 200240,
China}
\affiliation{Key Laboratory for Laser Plasma (Ministry of Education), IFSA Collaborative Innovation Center, Shanghai Jiao Tong University, Shanghai 200204, China}

\author{Boris A. Malomed}
\affiliation{Department of Physical Electronics, School of
Electrical Engineering, Faculty of Engineering, Tel Aviv University,
Tel Aviv 69978, Israel}

\author{Xianfeng Chen}
\affiliation{Department of Physics and Astronomy, Shanghai Jiao Tong University, Shanghai 200240,
China}
\affiliation{Key Laboratory for Laser Plasma (Ministry of Education), IFSA Collaborative Innovation Center, Shanghai Jiao Tong University, Shanghai 200204, China}

\author{Nicolae C. Panoiu}
\affiliation{Department of Electronic and Electrical Engineering, University College London,
Torrington Place, London WC1E7JE, United Kingdom}

\begin{abstract}
We demonstrate that graphene-based photonic superlattices provide a
versatile platform for electrical and all-optical control of photonic beams
with deep-subwavelength accuracy. Specifically, by inserting graphene sheets
into periodic metallo-dielectric structures one can design optical
superlattices that posses photonic Dirac points (DPs) at frequencies at
which the spatial average of the permittivity of the superlattice, $\bar{%
\varepsilon}$, vanishes. Similar to the well-known zero-$\bar{n}$ bandgaps,
we show that these zero-$\bar{\varepsilon}$ DPs are highly robust against
structural disorder. We also show that, by tuning the graphene permittivity
via the optical Kerr effect or electrical doping, one can induce a spectral
variation of the DP exceeding \SI{30}{\nano\meter}, at mid-IR and THz
frequencies. The implications of this wide tunability for the photonic
Zitterbewegung effect in a vicinity of the DP are explored too.
\end{abstract}

\pacs{Valid PACS appear here}
\maketitle

%
%
%
%
%
%

\affiliation{Department of Physics and Astronomy, Shanghai Jiao Tong University, Shanghai 200240,
China}
\affiliation{Key Laboratory for Laser Plasma (Ministry of Education),
IFSA Collaborative Innovation Center, Shanghai Jiao Tong University,
Shanghai 200204, China}

\affiliation{Department of Physics and Astronomy, Shanghai Jiao Tong University, Shanghai 200240,
China}
\affiliation{Key Laboratory for Laser Plasma (Ministry of Education),
IFSA Collaborative Innovation Center, Shanghai Jiao Tong University,
Shanghai 200204, China}

\affiliation{Department of Physical Electronics, School of
Electrical Engineering, Faculty of Engineering, Tel Aviv University,
Tel Aviv 69978, Israel}

\affiliation{Department of Physics and Astronomy, Shanghai Jiao Tong University, Shanghai 200240,
China}
\affiliation{Key Laboratory for Laser Plasma (Ministry of Education),
IFSA Collaborative Innovation Center, Shanghai Jiao Tong University,
Shanghai 200204, China}

\affiliation{Department of Electronic and Electrical Engineering, University College London,
Torrington Place, London WC1E7JE, United Kingdom}

\preprint{APS/123-QED}


\emph{Introduction.} One of the most daunting challenges one has to overcome
in order to develop ultra-fast active photonic nanodevices is the extremely
narrow tunability of physical parameters of optical media. For example,
while the conductivity of semiconductors can be varied by orders of
magnitude with the help of impurity doping or thermally, the index of
refraction of optical media, which is the main parameter defining their
optical properties, can hardly be tuned by more than \SI{0.5}{\percent}.
Several approaches that address this key problem have been proposed,
including the resort to plasmonic materials whose optical nonlinearity is
enhanced by strong optical fields associated with the resonant excitation of
surface-plasmon polaritons (SPPs) \cite{szg02prl,cvb11s,po04prl}, the use of
strongly anisotropic materials whose optical response can be controlled via
the polarization of the excitation field \cite{esz04ol,obt13prl},
reconfigurable metamaterials \cite{pak10nl}, embedding liquid crystals into
a photonic-crystal matrix \cite{wkk07oe}, and tuning the material refractive
index via carrier injection \cite{cpz06n} . While promising if targeted at
specific applications, the efficiency of these methods is generally limited
by high-power requirements, slow optical response, and reduced spectral
tunability.

A recently introduced photonic platform that contains all the ingredients
needed to achieve ultra-fast broad optical tunability at the nanoscale is
graphene \cite{Bonaccorso,Chen,Fei}. Compared to the SPPs in metals,
graphene plasmons are localized within much smaller regions, provide a much
larger optical near-field enhancement, and can pass significantly longer
distances \cite{Jablan,Koppens}. Equally important, the conductivity of
graphene and, implicitly, its electrical permittivity can be varied
substantially, locally and ultra-fast, by means of chemical doping or gate
voltage \cite{Yan,YanBis,Yao,Fang}. For example, changing the Fermi level by
a mere \SI{1}{\electronvolt} causes the relative variation of the real part
of the permittivity, at the \SI{10}{\micro\meter} wavelength, by %
\SI{250}{\percent}. Depositing graphene onto the surface of specific
materials, which may be both dielectrics and metals, a variety of
graphene-based layered nanostructures have been designed, and their
implementations in functional devices have been demonstrated \cite{Yan}. The
most salient feature of these photonic superlattices is that, by
incorporating graphene into their periodic structure, the band-structure and
thus their optical response can be dynamically tuned. This key property, in
conjunction with relatively low loss and extremely tight confinement of the
graphene plasmons, opens up promising applications of tunable graphene
photonic superlattices (GPSs) at mid-IR and at THz wavelengths. In addition
to these important practical considerations, GPSs can also play a major role
at the fundamental level. Specifically, it has been suggested that
electron-wave dynamics in graphene heterojunctions are in many ways similar
to the propagation of optical beams in photonic superlattices containing
positive- and negative-index materials \cite{Bliokh}, intriguing phenomena
such as the Klein tunneling \cite{Shytov,Young} and Zitterbewegung (ZB)
effect \cite{2DZB,1DZB1,Nam,1DZB2,Liang,1DZB3} being observed in both
physical systems.

\begin{figure}[t]
\center
\includegraphics[width=8cm]{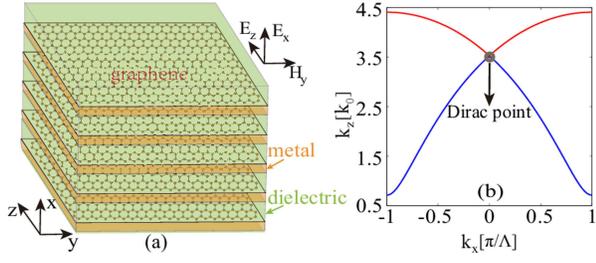}
\caption{(a) A schematic of the graphene photonic superlattice: the unit
cell consists of a graphene-metal-dielectric tri-layer configuration stacked
along the $x$ direction. The incident TM-polarized beam propagates along the
$z$-axis, its non-vanishing field components being $E_{x}$, $E_{z}$, and $%
H_{y}$. (b) The photonic band structure calculated in the
zero-averaged-permittivity case, featuring the 1D Dirac point. The structure
parameters are $\protect\varepsilon _{g}=-14.48$, $\protect\varepsilon _{%
\text{Si}}=12.25$, $\protect\varepsilon _{m}=-924.05$, $t_{g}=$~
\SI{0.5}{\nano\meter}, $t_{\text{Si}}=$~\SI{604}{\nano\meter}, $t_{m}=$~
\SI{8}{\nano\meter}, and $\protect\lambda =$~\SI{4.2}{\micro\meter}.}
\label{appa}
\end{figure}

In this Rapid Communication we demonstrate that, by including graphene
sheets into the unit cell of metallo-dielectric superlattices, one can
readily design GPSs that posses electrically and optically tunable Dirac
points (DPs) and feature the ZB of optical beams propagating near the DPs. A
remarkable property of these DPs is that they form when the spatial average
of the permittivity of the superlattice vanishes, $\bar{\varepsilon}=0$, an
immediate consequence of this fact being their remarkable robustness against
structural disorder. To add specificity to these ideas, we consider, as an
example of a GPS, the one-dimensional (1D) periodic structure depicted in
Fig.~\ref{appa}(a). It has a three-layer unit cell,
graphene-metal-dielectric, the metallic and dielectric layers chosen in this
work being silver and silicon, respectively. The dispersion relation of the
structure is easily found by using the transfer-matrix method (see
Supplemental Material \cite{SuplMat} for the derivation of this relation),
and is given by
\begin{align}
& \cos (k_{x}\Lambda )=\sigma \left( 1-\frac{\varrho _{ab}^{2}\varrho
_{ba}^{2}}{2\varrho _{ab}\varrho _{ba}}\tan \kappa _{a}\tan \kappa
_{b}-\right.  \notag \\
& \left. \frac{\varrho _{bc}^{2}+\varrho _{cb}^{2}}{2\varrho _{bc}\varrho
_{cb}}\tan \kappa _{b}\tan \kappa _{c}-\frac{\varrho _{ca}^{2}+\varrho
_{ac}^{2}}{2\varrho _{ca}\varrho _{ac}}\tan \kappa _{c}\tan \kappa
_{a}\right) ,
\end{align}%
where $k_{x}$ is the Bloch wave vector, $k_{z}$ the propagation wave vector,
$t_{a}$, $t_{b}$, and $t_{c}$ are the thickness of the layers, $\Lambda
=t_{a}+t_{b}+t_{c}$ is the period of the supercell, $\sigma =\cos \kappa
_{a}\cos \kappa _{b}\cos \kappa _{c}$, and $k_{j}=\sqrt{(\omega
/c)^{2}\varepsilon _{j}\mu _{j}-k_{z}^{2}}$, with $j=a,b,c$, where $a$, $b$,
and $c$ stand for graphene, silicon, and silver, respectively. For
convenience, we here define $\varrho _{ij}\equiv \kappa _{i}\tau _{j}$, $%
\kappa _{i}\equiv k_{i}t_{i}$, and $\tau _{i}\equiv \varepsilon _{i}t_{i}$.
By fixing frequency $\omega $ in Eq.~(1), dependence $k_{z}=k_{z}(k_{x})$
determines the spatial dispersion relation for the particular frequency.

\emph{Tunability and robustness of the Dirac points.} A peculiar property of
the band structure of such GPSs is that, if it possesses two transmission
bands, then the bands intersect at the center of the Brillouin zone, $%
k_{x}=0 $, giving rise to the photonic DP [see Fig.~1(b)]. For such a
two-band configuration to occur, the thickness of the dielectric and
metallic layers should be larger than certain critical values. As mentioned
above, a DP in our photonic structure exists provided that the real part of
the average permittivity of the superlattice vanishes, namely,
\begin{equation}
\mathfrak{Re}(\overline{\varepsilon })=\frac{1}{\Lambda }\mathfrak{Re}\left(
\tau _{g}+\tau _{m}+\tau _{d}\right) =0.
\end{equation}

DPs are well known in solid-state physics, being at the origin of many
remarkable properties of recently discovered materials, such as graphene and
topological insulators. In this case they are singular points in the
electron energy-band structure, and the aforementioned similarities between
electron and photon wave dynamics suggest that DPs can exist too in the
photonic band structure of certain periodic structures, one such example
being the GPS presented in Fig.~1(a). As we demonstrate in this study, the
graphene component of the present photonic superlattice allows one to tune
the wavelength of the photonic DPs by simply varying the chemical potential
of graphene, $\mu _{\text{c}}$. More specifically, changing $\mu _{\text{c}}$
by external means (e.g., gate voltage, chemical doping, etc.), one varies
the permittivity of graphene and therefore one can tune the wavelength, $%
\lambda _{D}$, at which condition $\mathfrak{Re}[\overline{\varepsilon }%
(\lambda _{D})]=0$ holds. These ideas are illustrated in Fig.~2(a), where we
plot the dependence of the wavelength of the DP on $\mu _{\text{c}}$. Note
that a spectral shift of more than \SI{30}{\nano\metre} can be achieved
changing $\mu _{c}$ by \SI{1.5}{\electronvolt}. The variation of the
corresponding band structure is presented in Fig.~2(b).
\begin{figure}[t]
\centering
\includegraphics[width=8 cm]{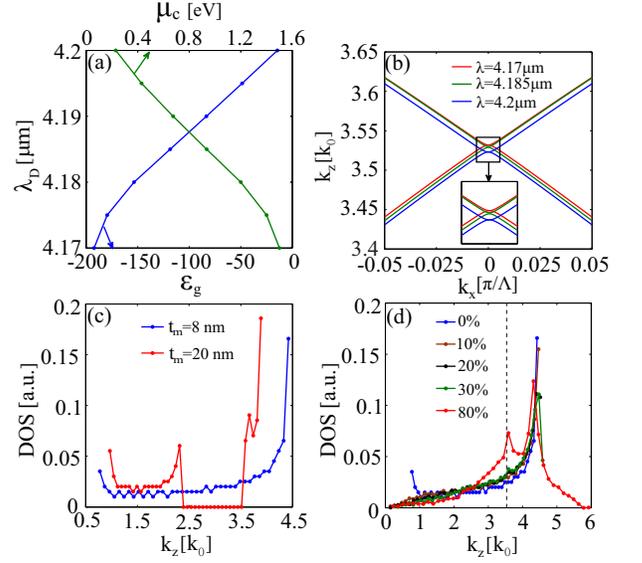}
\caption{(a) The dependence of the wavelength of the Dirac point on the
graphene's chemical potential and permittivity. (b) The photonic band
structure calculated for different wavelengths. Parameters are the same as
in Fig.~\protect\ref{appa}. (c) The density of states of a $50$-period GPS
possessing a Dirac point (at $t_{m}=$~\SI{8}{\nano\metre}) or a Bragg gap
(at $t_{m}=$~\SI{20}{\nano\metre}). (d) The density of states of the GPS in
(c) that possesses a Dirac point, calculated for different disorder levels.}
\label{appb}
\end{figure}

Similar to zero-$\bar{n}$ band-gaps, which are formed when the spatial
average of the refractive index of a Bragg grating vanishes, DPs in our
photonic structure are particularly robust against disorder added to the
system. This behavior is expected because structural disorder weakly affects
the spatial average of the permittivity, hence its vanishing, which
determines the existence of DPs, persists in the presence of disorder as
well. To illustrate this phenomenon, we present in Figs.~2(c) and 2(d) the
density of states (DOS) of periodic and disordered GPSs, respectively. In
the calculations, we used two superlattices, one selected so as to possess a
DP (at $t_{m}=8~\text{nm}$), whereas the other one is a regular superlattice
featuring a Bragg gap, at $t_{m}=20~\text{nm}$. The DOS of these two
superlattices is extracted from $dk_{z}/dk_{x}$, see the results in Fig.
2(c). As expected, while the DOS for the superlattice featuring a Bragg gap
exhibits a conspicuous wide region of forbidden states ($\mathrm{DOS}=0$),
the superlattice possessing a DP shows nonvanishing DOS in the whole
spectrum. The robustness of the DP against the structural disorder is
confirmed in Fig.~2(d), where we display DOS for the disordered superlattice
at several different levels of the disorder. It is introduced by assuming
random fluctuations of the metallic-layer thickness around its average
value, namely, the $n$-th thickness is $t_{m}^{n}=t_{m0}+\delta _{n}$ ,
where $t_{m0}$ is the average thickness (we choose $t_{m0}=%
\SI{8}{\nano\meter}$ here), and $\delta _{n}$ is a random perturbation. We
assume $\delta _{n}$ to be uniformly distributed in the interval of $%
[-\delta ,\delta ]$, hence the disorder level may be characterized by $%
\Delta \equiv \delta /t_{m0}$. The DOSs in Fig.~2(d) clearly demonstrates
that, even when the disorder level is increased to \SI{80}{\percent}, the DP
is preserved, appearing as a spectral peak in the DOS, at $\lambda
_{D}\simeq \SI{4.2}{\micro\meter}$.

Real-time tunability of the band-structure of the GPS provides an effective
way to control the excitation and propagation of optical beams. Figures~3(a)
and 3(b) show that, when one varies $\mu _{\text{c}}$ from %
\SI{0.23}{\electronvolt} to \SI{1.5}{\electronvolt}, the band structure
transforms from a gapless one, which possesses a DP, to a gapped structure,
this effect being accompanied by a dramatic change in the beam-propagation
dynamics. At the DP, a normally incident Gaussian beam splits into two, as
the upper- and lower-band excitations are not resolved, due to the fact that
the two bands have identical slopes near the DP [see Fig.~3(c)]. By
contrast, when a gap opens the output pattern exhibits four beams, due to
the difference in the slopes of the upper and lower bands near $k_{x}=0$ [as
per Fig.~3(d)]. Taking losses in graphene and metallic layers into account
yields similar patterns of the output optical field [Figs.~3(e) and 3(f)];
however, as expected, the output beams now decay in the course of the
propagation. In Figs.~3(e) and 3(f) we used the scattering time, $\tau =%
\SI{90}{\femto\second}$, which was measured in an ionic-liquid gating
configuration \cite{Ph}, the achieved Fermi level being as large as %
\SI{2}{\electronvolt} \cite{Ye,Efe}.
\begin{figure}[t]
\centering
\includegraphics[width=8 cm]{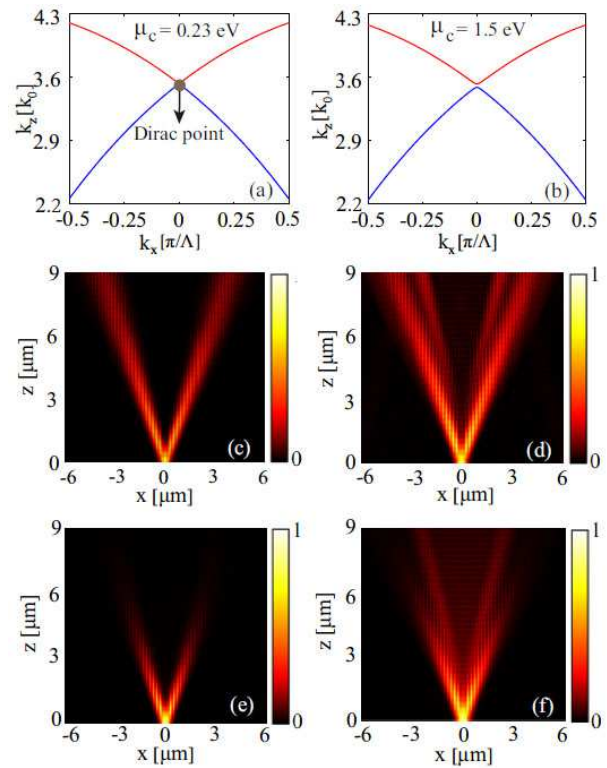}
\caption{(a), (b) Formation of a bandgap at the Dirac point with $\protect%
\lambda _{D}=\SI{4.2}{\micro\meter}$ by varying the graphene's Fermi level.
In (a) and (c), permittivities of the layers are the same as in Fig.~\protect
\ref{appa}. In (b) and (d), $\protect\varepsilon _{g}=-195.02$. The real
part of the permittivities in (e) and (f) are the same as those in (c) and
(d), respectively, whereas in (e) $\protect\varepsilon _{g}^{\prime \prime
}=1.28i$, $\protect\varepsilon _{m}^{\prime \prime }=56.37i$ and in (f) $%
\protect\varepsilon _{g}^{\prime \prime }=4.95i$, $\protect\varepsilon %
_{m}^{\prime \prime }=56.37i$. In all cases, the thickness of the layers are
the same as in Fig.~\protect\ref{appa}, and the input conditions select a
TM-polarized Gaussian beam with profile $E_{x}(x)=\exp (-x^{2}/w^{2})$,
where $w=2\protect\lambda _{D}$.}
\label{appc}
\end{figure}

\emph{Optically tunable Dirac points.} In addition to electrical means, the
wavelength at the DPs can be tuned by means of the intrinsic optical
nonlinearity of graphene, or of the dielectric and metallic layers as well.
Such nonlinearity-induced tunability of the DP is possible in our
superlattices because, as explained above, the DPs form when the real part
of the average permittivity is zero. Thus, nonlinear optical effects that
change the refractive index can be used to tune the permittivity in or out
of the zero-average state, thus providing all-optical means to control the
formation of DPs. The optical nonlinearity of graphene is expected to be
particularly strong, due to its unusually large third-order nonlinear
susceptibility \cite{Hendry}, as well as the tightly localized optical field
supported by graphene sheets.

Figure 4 illustrates the creation/destruction of a DP under the action of
the optical Kerr effect, induced in graphene and the neighboring silicon
layer. At low optical powers, starting with a superlattice for which $%
\overline{\varepsilon }(\lambda )<0$, the two transmission bands are
separated by a gap, as seen in Fig.~4(a). When the input power increases,
both graphene and silicon are affected by the Kerr nonlinearity, their
permittivities varying by $\delta \varepsilon _{g}>0$ and $\delta
\varepsilon _{\text{Si}}>0$, respectively. As a result, the average
permittivity of the structure, $\overline{\varepsilon }$, gradually
increases and the corresponding band structure is modified as well. At some
particular value of the optical power, $\overline{\varepsilon }$ becomes
equal to zero, so that, as shown in Fig.~4(b), the photonic gap closes and a
DP re-emerges. For the particular case illustrated in Fig.~4, $\delta
\varepsilon _{\text{Si}}=0.07$ and $\delta \varepsilon _{g}=0.46$. The
propagation of a TM-polarized Gaussian beam, both in the linear and
nonlinear regimes, is presented in Figs.~4(c) and 4(d), respectively. One
can see that, when the power of the input beam increases, the beam dynamics
changes from the typical discrete diffraction to 1D conical diffraction,
which is a manifestation of the formation of the DP in the latter case.
\begin{figure}[t]
\centering
\includegraphics[width=8 cm]{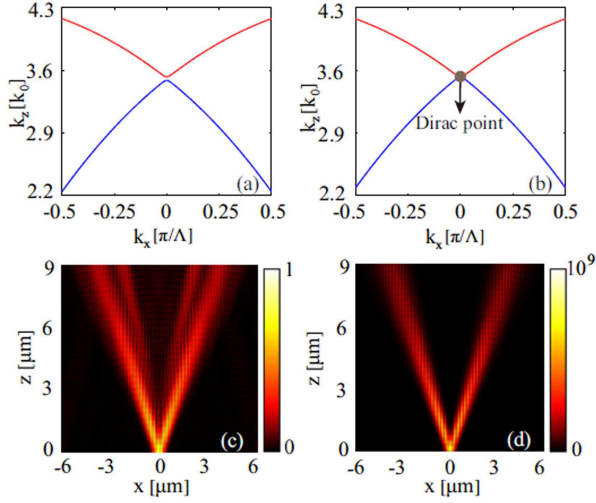}
\caption{(a) The linear band structure, calculated for the same parameters
as in Fig.~\protect\ref{appa}, except $\protect\varepsilon _{g}=-195.02$ ($%
\overline{\protect\varepsilon }=-0.15$). (b) The nonlinear band structure,
where the Bloch mode of the upper band at $k_{x}=0$ is used to induce the
nonlinear change of permittivities, $\protect\delta \protect\varepsilon _{%
\text{Si}}=0.07$ and $\protect\delta \protect\varepsilon _{g}=0.46$. In (c)
and (d), the input condition is a TM-polarized Gaussian beam with $%
E_{x}(x)=A\exp (-x^{2}/w^{2})$, with $w=2\Lambda $, and $A=1~\text{V/m}$ in
(c) or $A=4.13\times 10^{7}$~\textrm{V/m} in (d). The Kerr coefficients are $%
n_{2,\text{Si}}=1.86\times 10^{-20}~\mathrm{{m}^{2}/{V}^{2}}$ and $%
n_{2,g}=4.09\times 10^{-17}~\mathrm{{m}^{2}/{V}^{2}}$.}
\label{appd}
\end{figure}

\emph{Tunable Zitterbewegung effect.} Finally, we point out that the control
of the DP and band-structure of GPSs suggest a convenient way to implement,
in the classical-optics context, a series of intriguing quantum-mechanical
phenomena. As a relevant example, here we consider the ZB effect, which, in
the relativistic quantum mechanics, refers to rapid oscillations (trembling)
of a moving electron around its average position. The ZB of electrons arises
because of the interference of particle and antiparticle components of the
wave function. The electron ZB oscillations having extremely high
frequencies, of about \SI{e21}{\hertz}, therefore this phenomenon is not
accessible to experimental observations. Recently, optical analogs of the ZB
were studied in 1D \cite{1DZB1,Nam,1DZB2,Liang,1DZB3} and 2D \cite{2DZB}
photonic crystals. Here we demonstrate that GPSs are highly suitable
physical systems to study this effect. To this end, we show in Figs.~5(a)
and 5(b) the beam propagation at two different values of the chemical
potential, $\mu _{\text{c}}=\SI{0.23}{\electronvolt}$ and $\mu _{\text{c}}=%
\SI{1.5}{\electronvolt}$, respectively. Note that in both cases the GPS has
gaps in the wave-vector space, the gap at $\mu _{\text{c}}=%
\SI{1.5}{\electronvolt}$ being wider. In the latter case, the ZB has a
smaller oscillation amplitude and a larger frequency, which is consistent
with the findings reported in Ref.~\cite{1DZB2}. This behavior is more
clearly observed in Fig.~5(d), where we plot the evolution of the beam
center, $x_{c}\equiv \int_{-\infty }^{+\infty }|\mathbf{E}%
(x)|^{2}xdx/\int_{-\infty }^{+\infty }|\mathbf{E}(x)|^{2}dx$.

Still more important, GPSs provide a reliable platform to implement
\textquotedblleft nonlinear quantum mechanics", as shown in Fig.~5(c), where
we examine the ZB dynamics affected by the self-focusing nonlinearity,
associated with the graphene and silicon layers. The figure shows that,
while the ZB amplitude and frequency remain almost unaltered, the
self-focusing nonlinearity leads to a spatial walk-off of the ZB, \textit{viz%
}., monotonic spatial drift of the beam's center in one direction, in the
course of its propagation. Note that, as the thickness of the layers of the
GPSs can be scaled so that the real part of its averaged permittivity
remains equal to zero, one can easily extend the ZB effect, and the
emulation of other DP-related quantum phenomena, into the THz and other
spectral ranges, different from the standard optical one.
\begin{figure}[!t]
\centerline{\includegraphics[width=8.4cm]{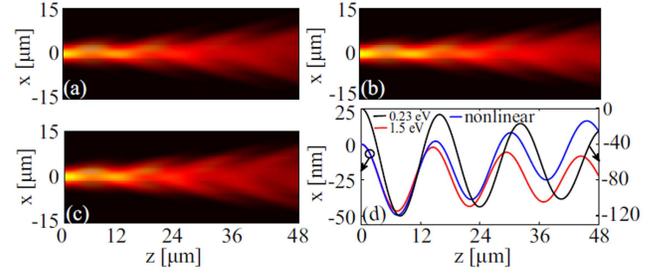}}
\caption{Beam propagation in the linear regime: (a) $\protect\mu_\text{c}%
=0.23~\text{eV}$, (b) $\protect\mu_\text{c}=1.5~\text{eV}$. (c) The
propagation in the nonlinear regime, with $\protect\mu_\text{c}=1.5~\text{eV}
$. (d) Dependence of the center of mass of the beam on $z$. The input
condition is a TM-polarized Gaussian beam with $E_x(x)=A \exp(-x^2/w^2)$,
with $w=12\Lambda$. In (c) $A=1.5\times 10^7~\text{V/m}$. The thicknesses of
the layers are $t_{g}=$~\SI{0.5}{\nano\meter}, $t_{\text{Si}}=$~
\SI{604}{\nano\meter}, and $t_{m}=$~\SI{10}{\nano\meter}, whereas $n_{2,%
\text{Si}}=1.86\times 10^{-20}~\mathrm{{m}^2/{V}^2}$ and $n_{2,g}=4.09\times
10^{-17}~\mathrm{{m}^2/{V}^2}$.}
\label{appe}
\end{figure}

\emph{Conclusion.} We have proposed graphene-based photonic superlattices
which should provide a highly effective and robust approach for the control
of photonic beams at the nanoscale. By taking advantage of the strong
dependence of the graphene permittivity on the chemical potential, it is
demonstrated that, due to the peculiar topology of their photonic band
structure, specially designed photonic superlattices, incorporating graphene
sheets interlaid with metallic and dielectric layers, feature large
electrical and optical tunability. Implications of our findings to
applications and studies of fundamental effects, such as Zitterbewegung, are
proposed too.

\emph{Acknowledgments.} The work of H. Deng and F. Ye was supported by
Innovation Program of Shanghai Municipal Education Commission (Grant No.
13ZZ022) and the National Natural Science Foundation of China (Grant No.
11104181 and 61475101). The work of N.C.P. was supported by the Engineering
and Physical Sciences Research Council, grant No EP/J018473/1. This author
acknowledges the hospitality of the Department of Physics and Astronomy of
Shanghai Jiao Tong University during a visit to which this paper was
written. B.A.M. appreciates a partial support through Grant No. B12024 from
\textquotedblleft Program of introducing talents of discipline to
universities" (China).

\end{document}